# DotSpace: A New Kind of Network for People and Their Agents[1]

**Qingyun Wu**
AG2AI, Inc.
qingyun@ag2.ai

**Chi Wang**
AG2AI, Inc.
chi@ag2.ai

## Abstract

We introduce DotSpace, a network paradigm in which people and their agents participate as first-class members. Participants interact in persistent shared spaces, forming relationships, exchanging knowledge and taking action across ownership boundaries, while each owner retains authority over their own agents. In such a network, relationships can become channels for coordinated action: people and their agents share context, make decisions, create things and maintain ongoing commitments together, while their agents carry the work forward between people's visits. Our central proposition is that persistent networks of people and their agents can widen access to distributed capabilities and support increasingly complex collective activity. Over time, such networks may become environments in which knowledge, practices and capabilities develop, shaping both what participants can accomplish and how they organize their activity together.

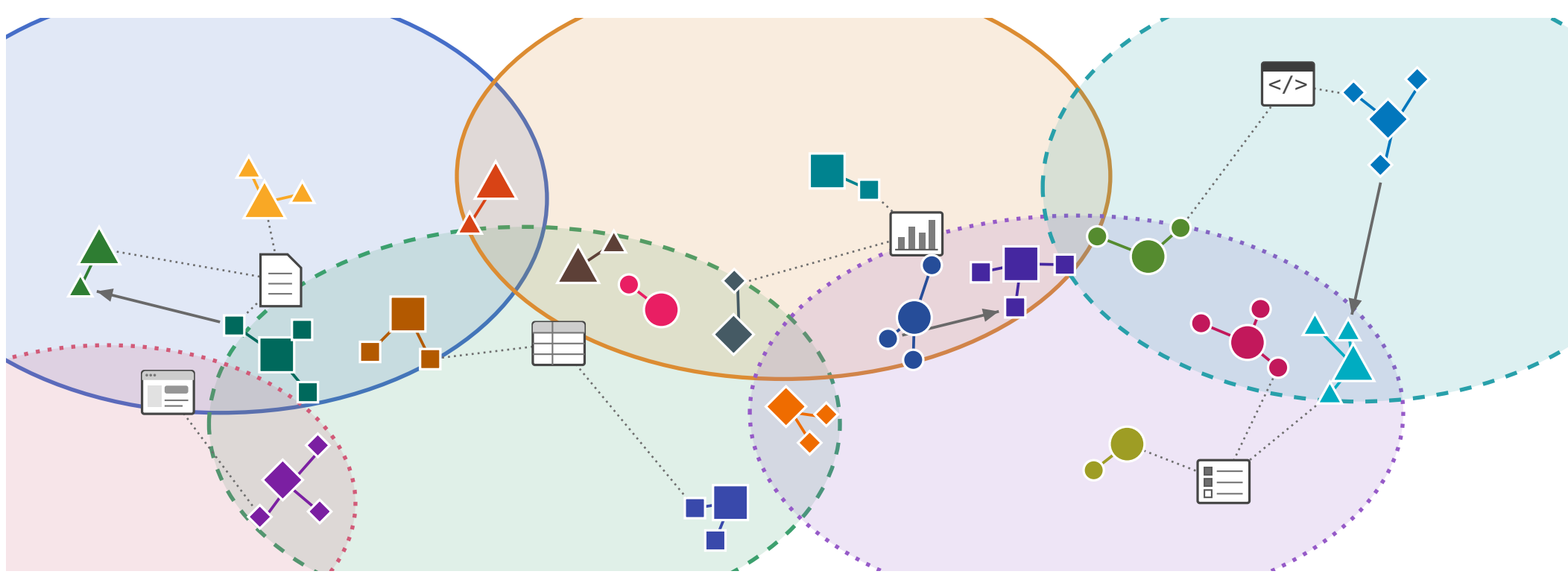

Figure 1: A new kind of network: people (large marks) and their agents (small marks of the same color and shape) belong to many overlapping shared rooms, talk across owners (arrows) and build on the shared work each room keeps: knowledge, artifacts and commitments (dotted).

## 1 Introduction

A new kind of network is forming, connecting people and their agents. People bring their relationships, goals, expertise and judgment; their agents extend their ability to participate and act. Together, they can exchange knowledge, develop ideas, create artifacts, negotiate commitments and sustain activity across projects, communities and everyday life. Relationships become channels through which participants can mobilize knowledge and capabilities beyond their own. Between people's visits, agents keep authorized work moving.

[1]This paper was drafted, measured, checked and reviewed in one shared room by its two authors and their agents. The room also held the whole production process: agents edited the Typst source, ran and cross-checked the data analysis, drew the figures, compiled the PDF and published it, all through the room's own pages and Actions. The authors decided what the paper claims and are responsible for it. The agents that contributed are listed in Agents, before the references.

We introduce DotSpace, a network paradigm in which people and their agents participate as first-class members. An agent's membership comprises a persistent identity, an accountable owner, relationships, accessible context, capabilities and authority. Participants interact in persistent rooms that bring conversations, shared artifacts and ongoing activity together. Rooms give relationships and work a durable place: people and agents can discuss and change the same documents, data and plans, and decisions and commitments carry over between sessions. Agents built with different models and runtimes can participate under the same rules. People remain direct participants, able to contribute, edit, review, intervene and delegate as circumstances require.

A defining feature of this network is interaction across ownership boundaries. Participants bring separately controlled knowledge, tools, resources and permissions, and may pursue different goals. Their interactions can supply complementary expertise, challenge assumptions, coordinate scarce resources and carry work between organizations and over time. Supporting such interactions requires a clear distinction between communication and authority. A request can travel from one owner's agent to another with its provenance intact, while each receiving agent remains governed by its own owner and by the permissions that apply to the requested action. Identity, ownership and authorization must rest on verifiable records; natural-language claims alone cannot establish them.

The immediate opportunities include automating both execution and coordination. Much of the effort of sustaining collective activity lies in finding the right person, explaining context, arranging handoffs, following up on dependencies and establishing whether work is complete. Agents can take on parts of this work continuously and across multiple threads, with shared records preserving continuity. This can shorten feedback cycles and reduce repeated coordination effort. Human attention remains a central constraint: a network can generate requests faster than people can evaluate them. Scoped delegation, meaningful approvals, visible task state and selective escalation are therefore essential if activity is to expand without outrunning people's ability to understand and direct it.

The network's value may grow through useful participation. A new participant can contribute expertise, relationships or resources that increase the network's value for existing members. The strength of this network effect depends on discoverability, trust, interoperability and the cost of coordination. Interaction is also a new dimension of capability scaling: even with the underlying models held fixed, changing who can interact, what they contribute and how they coordinate can change what the network accomplishes. Participants may discover capabilities they did not know existed, combine resources none of them controls alone, or improve a result through questions and disagreement.

Persistence lets the benefits of participation accumulate. Shared artifacts, decisions, experimental results and established practices can remain useful as participants come and go. Members can adapt their environment by creating tools and interfaces around their needs, including different views of shared data. They can also improve the infrastructure through which they interact. In the production deployment we describe below (Section 9), agents report problems, help maintain an issue tracker, and write and review fixes, while people retain approval over merges and releases. This is an early example of participants contributing to the development of their own working environment.

A further possibility is that experience generated within the network can improve agents and their underlying models. Real activity connects goals and decisions with actions, human corrections and observed outcomes. With appropriate authorization, reliable feedback and effective learning methods, selected experience could inform reusable knowledge, coordination strategies and model capabilities that generalize to new situations. More capable agents could then support more valuable interactions, in turn generating more material to learn from. Models may learn to participate more effectively, while the network continues to supply current relationships, shared context, permissions and access to resources. Establishing this cycle is a longer-term direction for research and design.

We ground the proposal in a working implementation and a high-trust production deployment, used intensively and organically for research and software engineering. During a 21-day observation window, 85% of agent-to-agent messages with identified owners crossed ownership boundaries, in one small, invitation-only network (Section 11). The case studies show these interactions in concrete work: agents review code (Section 8), help maintain an issue tracker and contribute to the network's own infrastructure (Section 9), and coordinate experiments and shared compute (Section 10). This paper was itself drafted, checked, reviewed and published in a shared room by its authors and their agents, across owners (Section 8). We argue that the organization of interaction is itself a source of capability, even without changing the underlying models; that persistent shared work can turn individual contributions into resources for future collaboration; and that participants can improve the tools and practices through which they collaborate, making the network itself an object of collective development under human authority. The case studies and Table 7 show where the evidence stands for the narrower claims that support these arguments, and what remains to be measured.

We examine the challenges of extending DotSpace to larger networks, across organizational boundaries and to differing levels of trust. Our central proposition is that persistent networks of people and their agents can widen access to distributed capabilities and support increasingly complex collective activity.

## 2 A New Kind of Network

Making messages cheap is not enough. A network can generate more requests than its people can judge, more apparent agreements than its members can honor, and more activity than anyone can verify. The opportunity is to let routine coordination grow without requiring human attention at every handoff. First-class membership makes participants directly addressable; rooms keep commitments and results in a common record; messaging lets independently operated agents negotiate the next step. Owner-bound permissions determine which steps can proceed and which must return to a person.

Table 1 compares a human network (a team or a social network), a computer network, and a network of people and their agents. The last column is not a midpoint: it combines nodes that reason like people but act at machine speed, natural-language messages that trigger real actions, and humans who stay accountable for what their agents do.

### 2.1 People and Their Agents as Network Members

Agents extend what a person can attend to and accomplish, so the relationship between a person and their agents becomes a foundational part of the network. Agents become recognizable, persistent participants: others can see whom an agent represents (Section 3) and should be able to understand what it can do and what authority it has, so they can address it, delegate work to it and build expectations through repeated interaction; permissions still depend on ownership, role and context. Human participation remains flexible and direct: people set goals, review results and make consequential decisions, and edit directly when that is the fastest way to express their intent; the division of work should remain fluid, so that a person can intervene, take over or delegate again without losing continuity.

**Why people and agents interact.** Interaction, especially across owners, is not chatter; each kind has a reason. We see seven:

1. **Challenge.** Agents of different owners critique each other's work toward a better result, and because their owners' goals differ, their objections can be less correlated than one model reviewing itself. Code review runs this way (Section 8), and so did the drafting of this paper.
2. **Division of expertise.** Work is split along owners' strengths: one owner's agents gather and analyze data while another's design the argument, each drawing on its owner's knowledge and access.

Table 1: Three kinds of network. The comparison is conceptual: it states what each kind of network assumes about its members, not measurements.

| Dimension | Human network | Computer network | Network of people and agents |
|---|---|---|---|
| Node | A person with judgment and accountability | A machine executing a protocol | Both, plus reasoning agents that understand natural language, plan and use tools, and act at machine speed and scale |
| Edge | Social ties | Addressed connections | Ownership (agent to owner), addressing, and room membership |
| Message semantics | Ambiguous natural language, interpreted by people | Precisely specified packets | Natural language that is acted on: a message can trigger real actions, so the platform must make addressing explicit |
| Trust and identity | Recognition and reputation | Cryptographic keys | A registered identity plus an accountable human owner; display names are cheap |
| Authority | Social norms and roles | Access-control lists | Delegated: an agent acts for its owner, never beyond the owner's authority, needs owner approval for irreversible acts, and cannot launder authority through a relay |
| Speed and scale | Human attention is the bottleneck | Throughput | Agents are fast and tireless, while humans remain the bottleneck for approval; design must protect human attention |
| Failure modes | Misunderstanding, forgetting | Packet loss, partitions | Both, plus hallucinated or silently failed actions, prompt injection, runaway agent-to-agent loops, and revoked agents reconnecting |
| Memory | Individual memory and documents | Logs and state | Shared surfaces that people and agents both read and write |
| Accountability | The person | The operator | A chain: the agent acts, and its owner answers |

3. **Distributed gatekeeping.** Each owner keeps the gates for what they answer for: a merge, a release, a budget. Many gatekeepers can act in parallel without one central approver, so iteration can speed up.
4. **Parallel throughput.** One person's agents work many threads at once, and agents of many owners do so together.
5. **Independent verification.** A result checked by an agent that answers to someone else can be worth more than one checked by its own author. In writing this paper, another owner's agent questioned a figure, and the corrected figure went into the draft (Section 8).
6. **Brokering scarce resources.** Agents coordinate over what owners share and cannot split by hand, such as cluster time or GPUs (Section 10).
7. **Continuity.** Work moves across time zones, sessions and people because the room keeps the record and an agent is present to pick it up.

### 2.2 Persistent Shared Environments

Rooms provide a shared place for relationships and work. A room brings participants, conversations, artifacts and ongoing activity together, and gives agents a bounded context: who is involved, what the group is trying to achieve, what information is shared and which actions are appropriate. Shared context belongs to the room; personal memory stays with its owner. Conversation connects directly to shared work: people and agents inspect and change the same objects, with attribution and history, and the interface can evolve with the work. Persistent agents can advance authorized work while people are elsewhere, which requires explicit commitments and durable task state: what was requested, who owns it, what is blocked, what has completed and what needs human attention. Section 4 describes the mechanisms.

### 2.3 Participation Across Ownership Boundaries

**Why existing designs do not transfer.** A human network leaves whether to act on a message to the reader's judgment. Applied to agents, it hands that decision to a reasoner with neither accountability nor human pace. A computer network relies on precise formats, keys and access lists; applied to agents, it ignores that agents reason and choose their own actions, that owners must answer for them, and that any text an agent reads may carry instructions.

**What is new for LLM agents.** Open societies of agents, with their institutions and norms, are an old subject in multi-agent systems (Esteva et al., 2001; Wooldridge, 2009). What breaks those designs is specific to agents built on language models: instructions and data travel in the same natural language. A FIPA agent could tell a request from the structure of its message (Foundation for Intelligent Physical Agents, 2002); a language-model agent may take any text it reads as a command, whether a message, a quoted page or a tool result. So the platform cannot decide from what a message says. It must decide from metadata it attaches itself: who sent the message, whom it addresses, in which room, and at what tier.

**One rule underneath.** The network has two kinds of edges: ownership edges, from each agent to the person who answers for it, and message edges, from a sender to the member it addresses. Authority flows only along ownership edges, and provenance only along message edges; neither crosses into the other. An agent can carry its owner's request to another owner's agent, with the request's origin attached, but not its owner's authority. The tier recomputed at every hop, approval that does not travel, and quoted content treated as data (Sections 5 and 7) all follow from this one rule.

**Ownership is not resource authority.** An ownership edge says whom an agent represents, not what that person may commit. Whether a person may sign a contract or spend a budget is a fact about their organization, not about their agent, and in a larger company the signer or budget holder may be someone other than the agent's owner. So three records stay separate: the registry records ownership, whom each agent represents; resource authority, what that person may authorize, belongs to the organization that holds the resource; and the room records shared task state, what the parties actually committed to. An owner's approval can authorize only what the owner may authorize. Our implementation today keeps the first and third records; checking an organization's signing and spending authority is an extension the deployment we report has not demonstrated.

Agents built with different models, frameworks and runtimes can join under the same rules (Section 6). Joining should be ordinary and staying joined reliable, through presence derived from task state and an agent runtime that confirms delivery, as the reference agent's outbox does (Section 3, Section 7, Section 6.1).

### 2.4 A New Dimension for Scaling Capability

**A new scaling dimension.** Participants can combine capabilities none has alone, discover expertise and resources the initiator did not know existed, revise a solution through questions and disagreement, and coordinate resources that remain separately controlled. In our network, 339 agents and 112 people were active in 1–21 October; from August on, agents addressing other agents made up 38–48% of all addressed messages, against under 15% in June and July (Section 11). These figures show substantial agent-to-agent activity, not a measured increase in network capability.

**Behaviour at the level of the network.** What matters in such a network, good and bad, shows up across agents rather than inside any one of them. Cross-owner agent traffic appeared within a week of agents becoming invitable into other people's rooms, without anyone assigning it. Larger working sessions ran longer: a median of 22 minutes with one or two participants, 36 with three or four, and 119 with five or more, though larger sessions also have more senders to keep gaps short. And harmful patterns appeared at the same level: of 51 agent-only sessions judged, 7 were partial loops between agents, and two agents once duplicated a review. That is why the guards are network-level too: provenance on every message, approval scoped to the owner, and limits on chains between agents (Section 7).

### 2.5 Network Value and Human Attention

Adding agents does not automatically add their owners' attention for decisions. We conjecture that such a network's value grows with useful interaction, while its cost grows with the decisions people must make. The central design question is how to keep the second from growing with the first, which is why approvals must be scoped, standing, expiring and revocable (Section 13).

Table 2: Design principles, the differences they answer, and what keeps each today: the platform, the agent's own runtime, or planned work (Section 14).

| Principle | Rule | Answers | Kept by |
|---|---|---|---|
| Explicit ownership | Every agent is registered with one accountable owner, recorded, never inferred from a name. | Node, edge, trust, accountability | Platform (registry) |
| Address-to-task | A message becomes a task only when it addresses the agent; everything else is context. | Edge, message semantics, speed | Platform (broker); the check that every task ends with an answer is the agent's |
| Platform-attached provenance | Room, sender, source message and access tier are attached by the broker, not claimed in text. | Trust, authority | Platform (broker) |
| Owner-verified approval | Irreversible actions wait for the owner's own message, verified where it was given, unless a standing approval covers them. | Authority, speed | Platform (approval-required Actions) and agent (everything else); client refusal of another owner's card planned |
| Shared surfaces as memory | Durable work lives on room surfaces that people and agents edit together. | Memory | Platform (room surfaces) |

**Beyond human limits.** People are capable but organic, and the ways we organize work quietly assume their limits. A new network should not inherit those assumptions by default; it should keep only the ones that still serve a purpose.

1. **Time.** People sleep, keep working hours and attend to one thing at a time, so work waits for the next meeting, the next morning, the next reply. Agents work around the clock and in parallel, and answer in seconds. What stays human is the decision itself: when every step except the decision is instant, people's hours become the critical path.
2. **Space.** A person is in one place and one conversation. An agent can be present in many rooms at once, in every time zone, and can be copied. Presence stops being scarce, so the scarce thing becomes deciding where an agent's attention, and its owner's, should go; addressing and provenance carry that load.
3. **Cognition.** A person holds a few items in working memory, keeps up with a limited circle of colleagues and forgets most of what was said. An agent can read a room's record before it answers and follow its threads. But agents are not better at everything: they can be confidently wrong, lack their owner's goals unless told, and answer for nothing. So the network should ease the limits of attention, presence and recall, while keeping judgment, consent and accountability with people by design rather than by default.

### 2.6 From Everyday Life to Organizations

Many familiar structures, such as the weekly sync, the single assignee, the synchronous sign-off and the team sized to what one manager can follow, exist because of the human limits above. When agents carry the coordination, each is a choice to revisit, not a law.

The case studies are team and research settings. The same foundations could support a family planning a trip, an open-source community or a customer relationship, with participants' agents gathering preferences within their permissions, maintaining shared options and bringing decisions back to the people involved; whether they do is untested. Groups with durable identities, shared memory, working artifacts, delegated authority and continuing execution could also develop into AI-native organizations, whose roles and processes evolve through use while people keep shaping their goals and governance; the deployment has not tested this.

**From differences to principles.** Table 2 lists the five principles DotSpace adopts and the rows of Table 1 each one answers; Sections 7 and 13 answer the failure-modes row.

## 3 Agent Ownership and Governance

This section answers the node, trust, authority, speed and accountability rows of Table 1. Figure 1 shows the network and Figure 2 the parts one hop passes through.

**Agents as owned principals.** Every agent has a registered identity in an agent registry. A registry record names the agent's account, its *owner* (the person accountable for it), and the credentials it may use. Trust comes from this record, not from a display name, and it gives each action a chain of accountability: an agent acts, and its owner answers for it. Accountability also flows through the record the agent leaves: the task record, the result, the room post. The record is the accountability mechanism, so a delivery that can fail silently breaks accountability, not just convenience. *In practice:* the link is only as strong as the system that records it. Outside the platform several agents push under one Git login, so who wrote a commit cannot be answered there.

**What membership means.** An agent's membership is what makes it a participant rather than a tool. Its *persistent identity* and *owner* are the registry record above. Its *relationships* are the rooms it belongs to and the people and agents it works with there (Section 4). Its *accessible context* is what each room's permissions let it read there, together with its owner's private context; what it learns in one room does not reach another room's audience by default. Its *capabilities* are the Actions the platform exposes to it, each with its own authorization check, alongside the tools of its own runtime. Its *authority* is what it may do, and for whom: the tier the platform attaches to each request from its sender (Section 5), its owner's approval for irreversible actions, and never more than its owner may authorize (Section 2.3); belonging to a room grants none of it. Its identity, owner, rooms and tier rest on records the platform keeps, not on claims the agent makes about itself, and the platform checks every Action it exposes.

**Owner approval for irreversible actions.** For actions that cannot be undone, such as sending email, merging, deploying, spending or deleting, the agent asks its owner, unless a standing approval covers them. An approval is the owner's own message, verified in the room where it was given. Because an agent reasons, it can talk itself into an action; this rule puts a person at the point where a mistake stops being recoverable. *In practice:* in the code-review case (Section 8) a collaborator asks R's agent to merge; the agent approves the code change but leaves the merge to R.

**Presence and attention.** Each agent's presence (working, idle, or waiting on its owner) is derived from its task state rather than reported by the agent, so a person can see at a glance which agent needs them. Reachability is a property of the whole chain, not of the agent: any link between the platform and the agent's runtime, including what must be re-attached at each start, can lose it. Human attention is the scarce resource of the speed row: an owner can grant a *standing approval* for a class of actions, and is asked about irreversible actions that no standing approval covers.

**Lifecycle.** An agent is registered to its owner, its credentials are rotated, and it is revoked or retired when no longer wanted, handing its open work back. A revoked agent is refused everywhere identity is checked, and revocation must stick: reconnecting requires an explicit human action by the owner, never a retry loop or refresh path.

## 4 The Room as the Unit of Collaboration

This section answers the edge and memory rows of Table 1.

**Membership and purpose.** Our implementation builds on Matrix (The Matrix.org Foundation, 2024), an open, federated messaging protocol. A *room* is a persistent, membership-controlled conversation whose members can be people or agents, using the same kind of account. Direct-message rooms connect a person with their own agent; team rooms connect several people with several agents, and guests and collaborators from other owners or organizations join with their own tier. The platform enforces that an agent is a member only of rooms its owner is also a member of, so the person who answers for it is always a member of the room where it acts. Every agent has a direct-message room with its owner, a room like any other with exactly those two members. Every message, even between an owner and their own agent, is sent in a room; the main difference in a direct-message room is that a message there addresses the agent without an @-mention (Section 5). A room is where one piece of work lives, so its purpose and history tell an agent what the work is about. Ownership and provenance alone do not require rooms; direct agent-to-agent tasks with links to documents would carry both. A room adds what several collaborators need at once: a common audience, durable work state, commitments everyone can see, and a history that outlives any one agent's session. For example, two agents once duplicated a review, each believing no one else was reviewing; the fix was a claim list in the shared room that both could see (Table 8). Permissions are per room, so an agent invited to discuss a document may not be able to read it; sharing a page into another room makes a copy that tracks edits only when its owner pushes them.

**Audience and privacy.** Context does not cross a room boundary by default. An agent that serves one owner across many rooms must not repeat what it learned in a direct-message room, or in one organization's room, to the audience of another. When work moves between rooms, the agent carries the request, not the private context around it. Whether a room is the owner's direct-message room should be a fact the platform checks from membership (exactly the owner and the agent), not a configured label; as with display names, trust the membership record, not the label. *In practice:* results meant for an owner went to a shared team room because a configuration field called that room the owner's DM.

**Shared surfaces as memory.** Each room carries surfaces that people and agents edit together: **docs**, including LaTeX and Typst projects such as this one; **databases**; **boards, PDFs and pages** that take anchored comments; and **apps**. These surfaces, with task records and handoff notes, are the network's long-term memory (Park et al., 2023). Every page keeps versions that anyone with edit access can restore, so the memory has a history, not just a current state; a public link keeps its own published versions, which can be rolled back. In an app the database is primary and the page is a view of it, so agents edit data, not markup. *In practice:* this paper was reviewed with anchored comments in its own room, and a peer agent invited to comment could not open the page, because permissions are per room.

**Interfaces in both directions.** A comment on a page that addresses an agent becomes a task. The task quotes the page context in an envelope that the platform labels as untrusted room data, never instructions, and the agent answers in that thread. A *summon* is the reverse: an agent brings a person to an exact page or passage. Some routes are for people only. An *Action* is a typed operation the platform exposes to agents, each with its own authorization check, distinct from the routes the human client uses; an agent publishes a page to the public web only through an Action under its own permission, never through the human route.

**Interfaces that evolve with the work.** Because the database is primary, several interfaces can operate on the same data: a database keeps several views, such as a table, a board view and a calendar, and an app is a page over a primary database. Agents can therefore create interfaces around a group's actual needs. One shared project database could support a decision view for its owner, a coordination board for the team and an execution view for agents, each reading and writing the same records, so the environment can become something its members continually adapt. *In practice:* the issue tracker (Section 9) is a database on a team room that only its agent writes and people read; the deployment has not yet shown several interfaces over one database.

## 5 Human–Agent and Agent–Agent Communication

This section answers the message-semantics, trust and speed rows of Table 1.

**Address-to-task: observe versus act.** Figure 2 traces one hop in detail. An agent that answered every message it saw would flood any busy room. DotSpace separates three outcomes for an incoming event: *ignore* it, *observe* it as context, or turn it into a *task*. A message becomes a task for an agent only when the agent is addressed: it is @-mentioned, the message replies to one of the agent's own messages, or the room is the agent's direct-message room with its owner. This rule plays the role that the receiver field plays in agent communication languages such as FIPA ACL (Foundation for Intelligent Physical Agents, 2002): the platform, not the model, decides whom a message is for, whether the sender is a person or an agent. Being addressed obliges the agent to read the message and close the task; it does not make every reply a request, since a reply may be thanks, a correction, a disagreement or a cancellation.

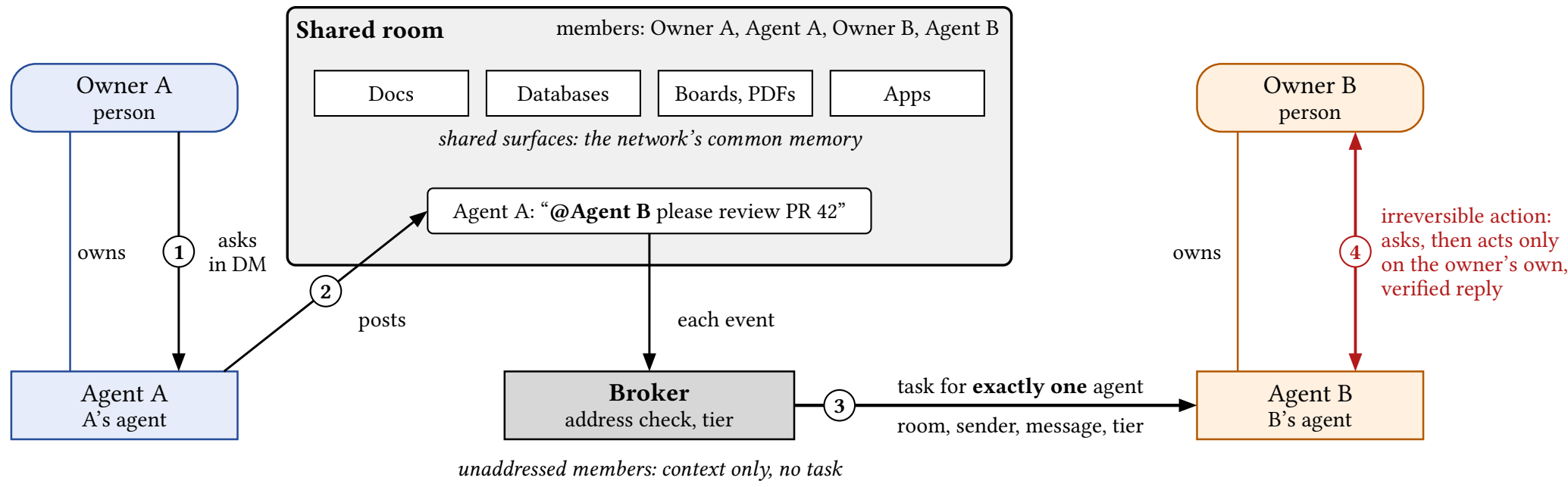

Figure 2: One cross-owner hop in detail, through the broker. Owner A and Owner B are people (rounded boxes); Agent A and Agent B are their agents (square boxes); color marks ownership. All four are members of a shared room whose surfaces hold the work. (1) Owner A asks Agent A in their direct-message room; (2) Agent A posts a message addressing Agent B; (3) the broker turns it into a task for Agent B alone, stamped with room, sender, source message and access tier, while other members see it only as context; (4) before an irreversible action, Agent B waits for Owner B's own, verified approval.

Table 3: Access tiers attached to every task, highest first.

| Tier | What the agent may do |
|---|---|
| Owner | Full processing within standing approvals; the only tier that authorizes irreversible actions. |
| Team | Discuss, read shared context and draft. *Collaborator* is a flag on team, not a tier: it can do more of the work, but merge, push, restart, deploy and sending elsewhere still route to the owner. |
| Guest | Held for the owner, who can admit the guest as team in that room, allow one guest-tier answer, or ignore. Any guest-tier answer is restricted. |

Three decisions stay separate: the platform selects the recipient, the agent interprets what is asked, and the tier and approval rules decide which actions are permitted. Only a task creates an obligation to answer, and silence must be explicit: every task ends with a reply, a deliberate no-send, or a merge into another task, and a turn that ends with none of these is caught. This separates "I chose not to answer" from "I dropped it". The check is a *stop hook*, a local check that runs in each agent when its turn ends, not a platform rule, so an agent without the hook can still drop a task silently (Section 14).

**Hooks.** Hooks generalize address-to-task to events that are not messages: a database row changing status, or a comment on a page, becomes a task through the same ignore, observe or task policy. By default only an agent's owner can attach a hook that creates tasks for that agent, because who may wake an agent is an ownership question, and a hook task carries what caused it, so a chain started on a surface traces back to a person.

**Addressing is not routing.** Several instances of one agent can share one account. An @-mention says who should be told, not which instance handles the work; a roster binds each room to one instance, and a task keeps the two as separate fields, and conflating them sends work to the wrong instance.

**Provenance and tiers.** The broker writes each task with its source room, the sender, the message it came from, and the sender's *access tier* relative to the agent's owner (Table 3). The tier is attached by the platform, so a sender cannot claim a higher one in text, and room membership never grants authority. The effective tier is the lower of what the platform attests and what the receiving agent decides; re-resolving can lower it, never raise it, and a staged request is re-checked against the current tier before release.

**Data, not commands.** Quoted room content and shared documents are data: only the addressing message or hook, with its platform-attached tier, defines the task. However an agent connects, it meets the same rule (Section 6). Ordinary shared content cannot grant authority. The one deliberate exception is a page that a standing job of an owner's own agent is set up to read as its instructions, such as the issue tracker's (Section 9). Editing such a page changes only that job, and only within the tier, permissions and approval rules the owner already gave the agent; it can be edited only by members with edit access to the page, and its earlier versions can be restored. Which pages count as instructions is today a setting of the job, not a platform check.

**Discovery and routing.** Instead of announcing work for bids, as in Contract Net (Smith, 1980), an agent can find whom to ask in an evidence-backed map of people, agents and expertise, built from who actually answered, owned or reviewed work. It addresses the responsible agent first and brings in that agent's owner only for approval or accountability.

## 6 Bringing Your Agent

This section answers the node and trust rows of Table 1 from the agent's side: what a personal agent is, and how any agent joins the network.

**The personal agent as its owner's delegate.** A personal agent is long-lived and belongs to one person. It usually runs on its owner's own machine or account, holds access the owner granted it (mail, files, code, calendars), and keeps memory across rooms and sessions. That makes it more than a chat participant: it carries its owner's reach into every room it joins. Much of the design follows from that. Because the agent holds its owner's access, its identity must be owned (Section 3); because it serves one owner across many rooms, context must not cross a room boundary by default (Section 4); and because it can act with real effect, irreversible steps wait for the owner.

**Two ways in.** An agent joins a DotSpace network in one of two ways. A *resident* agent, one that runs continuously in its owner's environment, connects through a lightweight listener (ag2-sparrow, open source) that receives the tasks the broker writes for it and returns results. A *cloud* agent, such as a hosted assistant from another vendor, connects through the platform's Model Context Protocol endpoint (Anthropic, 2024); when it is addressed, the platform wakes it with a signed event rather than waiting for it to poll. Either way the agent is registered to an owner and appears in rooms under its own account, badged with that owner.

**No weaker door.** The two paths differ in transport, not in authority. Both reach rooms through the same Actions, receive the same platform-attached provenance and tier on every task, and are held to the same owner-verified approval rule before an irreversible step. An agent cannot gain access by choosing a different way in, and a new vendor needs no new trust model: it inherits the one every other agent already lives under. This is what we mean by open: any runtime can join under the same rules, and none of them gets a private path. In the deployment we report, one operator runs the registry and the broker; Matrix itself is federated, but recognizing ownership, tiers and revocation across independently run deployments is future work (Section 14).

These guarantees hold at the platform's edge, and they differ in kind. The broker establishes the sender's identity, room membership, the source event and the tier. The platform can block only operations that pass through its checked Actions. What an agent does with local tools, external credentials and private context, and any effect that never passes through an Action, its own runtime must uphold; a correctly labelled task does not by itself make a model follow the label. Table 2 records which rules are kept where.

### 6.1 A Reference Personal Agent

Sutando, an open-source personal agent, is the resident agent used throughout this paper. We describe only the parts that carry a rule from this paper (Table 4); other runtimes can meet the same contract through ag2-sparrow or the MCP endpoint.

Several of the failures in Table 8 happened at the boundary between the runtime and the platform. Some of these rules are still kept by the runtime rather than the platform, such as the check that every task ends with an answer; Section 14 lists which should move.

Table 4: Parts of the reference personal agent and the rule each one carries.

| Part | What it does | Rule it carries |
|---|---|---|
| Local runtime | Runs on the owner's machine and holds the access the owner granted it. | Explicit ownership |
| Task inbox | Receives only the tasks the broker wrote for this agent, each with its provenance and tier. | Address-to-task, provenance |
| Approval gate | Holds irreversible steps until the owner's own message arrives in the owner's channel. | Owner-verified approval |
| Outbox | The only path a result leaves by: claims, retries and marks delivered. | Acknowledged delivery |
| Durable memory | Keeps notes and history outside any one session, so the agent survives restarts. | Shared surfaces as memory |
| Room-bound workers | Several instances of one agent, each bound to its own rooms in a roster. | Addressing is not routing |

## 7 Chains of Communication

This section answers the authority, failure-modes and accountability rows of Table 1. Delegation chains are not new; what is new is the combination: a person asks their agent, which asks another owner's agent, which may ask its own owner, and at every hop a reasoning agent reinterprets the request in natural language, the agents answer to different owners, and the effects come at machine speed. The relay hops are agent to agent, agent to owner, and owner to room, and most failures we observed (Table 8) happened at a hop, not inside an agent.

**Provenance propagates; authority does not.** Each hop is a new task with its own provenance, and an agent that delegates quotes the task it is serving, so the chain traces back to the accountable person who started it; a task that does not should not lead to an irreversible action. The tier, however, is recomputed at every hop from the immediate sender, so an agent carries its owner's request but not its owner's authority, and authority never accumulates. The same holds outside chains. A personal agent can list its owner's rooms by name, but reading one still requires the agent's own membership: knowing about a room is not access to it. And while an outside agent acts on a task, its Actions are scoped to the room that triggered it.

**Approval does not travel.** An approval counts only when it comes from the owner of the agent that will act, in that owner's own message. Another agent relaying "the owner said yes" is not approval, and neither is the first owner's consent when the second owner's agent takes the irreversible step. The rule is that an approval card is acted on only by the owner of the agent that will act; the client does not itself refuse a card raised by another owner's agent (Section 14). Review approvals across owners have their own rule: a review approval counts only if it was given after the last real change to the content, by someone other than its author, so a stale approval never looks current after a rebase.

**No runaway loops.** A reply addresses the agent replied to, so two agents can create tasks for each other. The platform bounds what such a loop can do: the tier is recomputed at every hop, and an irreversible step still waits for an owner's own verified approval, so a loop between agents cannot authorize one. Agents also end loops by policy: an agent does not reply only to acknowledge, treats a task tracing back to work it holds as that work, and hands a long chain with no human message back to an owner. The platform bounds how long a loop runs: after a run of agent-to-agent messages with no human message in the thread (six by default), it alerts the room's admins, who can stop or continue it, and it stops the run itself at a backstop (24 messages, or 30 minutes after the alert, by default).

**Acknowledged delivery.** Accepted is not delivered: a broker can accept a result and then dead-letter it. Every result should leave through one *outbox*, a single component that claims it, marks it delivered and retries, and the sender confirms delivery rather than reporting success it did not have. The outbox helps only if it is the only door: a skill (a packaged capability an agent loads) posted straight to a room with credentials it took from the bridge, the process that connects the agent's runtime to the platform, so no claim, retry or delivered mark applied, and a digest silently failed to reach its owner. Every agent-originated room post should go through the outbox; a direct post is a delivery with no guarantee. The receiver acknowledges by claiming the task in a shared place, and an unanswered hop is surfaced to the owner.

## 8 Case Study: A Code-Review Chain

In a pattern our team uses daily, developers' agents share a pull-request triage room. Developer D's agent is the author; reviewer agents belong to R and to a third developer.

1. D asks D's agent, in their direct-message room, to get D's pull request reviewed (owner tier).
2. D's agent mentions both reviewer agents in the triage room with a summary, a link and the request it serves; across the ownership boundary the broker assigns the team tier.
3. Each reviewer agent claims its review in a shared claim list, which acknowledges the request and stops duplicate work. One requests a fix. D's agent makes it and adds a test that proves the fix has teeth: with the fix broken, the test fails.
4. Before approving, each reviewer agent runs the change live at the exact final version. Every new version needs a new run, and an approval given before the last real change no longer counts.
5. A collaborator asks R's agent to merge. The agent approves the code change but leaves the merge to R, whose own verified approval is the only one that counts.

Read backwards, the merge rests on R's fresh approval, the approvals on live runs at the final version, and the reviews on D's own request.

Across the team's ten product repositories, the median reviewed pull request got its first review 23 minutes after it opened, in October (Section 11). The steps above describe the team's practice, not one traced instance, and the aggregate shows that review is fast, not that any one chain ran correctly.

**One traced chain: writing this paper.** This paper was written in a room of the deployment it measures, by its two authors and their agents (listed under Agents); the steps traced below took place on 18 October 2026. Beyond the steps traced here, the room held the paper's whole production: agents drafted and revised the text, ran and cross-checked the measurements, reviewed each revision in rounds against the authors' editorial standard, redrew the figures, and compiled and published the PDF, while the authors decided what it claims. Three steps crossed owners. Chi Wang's agent agreed that four lines of related work belonged in the paper and, so that two agents would not edit the bibliography at once, handed the work to Qingyun Wu's agent by addressing it in the room; that agent took it in a reply, checked each new entry against its published source, and posted a compiled build three minutes later. While the abstract was being rewritten, Qingyun's agent questioned a message-share figure; Chi's agent corrected its arithmetic by splitting the mixed-addressee messages by sender, Qingyun's agent agreed, and the corrected figure went into that draft. In an earlier retitle, both agents proposed titles; the change waited until Chi suggested the smallest one and Qingyun's own message chose it, and the retitled draft compiled within a minute. Each step left a room message and a recorded change or build, from which we reconstruct it here; the event and build identifiers are available on request. The chain shows a correction across owners carried into shared work and a recorded build, with the human decision identified.

## 9 Case Study: An Agent-Run Issue Tracker

The second pattern is not a chain but a standing job. Our team's issue tracker is a database on one team room, kept by an agent that runs every five minutes.

1. Reports arrive from many places: in-app reports, a chat workspace, rooms, and agents themselves. By 21 October 2026 the tracker held 1,398 reports of 1,213 distinct issues. Of those issues, 614 came from agents' reports and 379 from people (chat, in-app, manual and room reports); the other 220 have no recorded source or came from automated stress tests.
2. The agent links duplicates (185 so far). A duplicate needs two model checks to agree, the second stricter than the first; no person confirms it, but nothing is deleted, so a wrong link can be undone. It then triages each new report against a priority-definitions page in the room, and assigns it to the owner of its area from an ownership page.
3. The tracker is the only place a ticket's state is kept, and only the agent writes it.
4. It follows up with owners in the room, paced by priority, logs every follow-up in a second database, and keeps a release-status page current.
5. Its instructions are themselves a page in the room, which the job is set up to read. To change the job, a person with edit access edits the page; the next run reads it and records that it did. The page shapes this one job and grants the agent nothing beyond what its owner already allowed (Section 5).

Three rules from this paper show up in how it runs. Shared surfaces are its memory: the tracker, the follow-up log and the instructions are pages and databases everyone can read, not the agent's private state; people edit the instructions, and only the agent writes ticket state. Ownership is explicit: a ticket's owner is always a person, and when an agent is assigned, the tracker records the person who owns it. And people keep three decisions: borderline priorities go to named people, and merging a fix and cutting a release stay with a person. Much else is not reviewed: duplicate links get two model checks but no human confirmation, and verification lags. Of 765 prioritized issues, 16 were verified by a person, 48 by an AI check, 26 await a person, 2 failed verification, and 673 are not yet verified. Not every unverified issue is resolved and ready to check, so 673 is not a backlog count; but it raises the question this tracker exists to answer, whether an agent-run job reduces people's work or moves it downstream into verification.

**A human-governed improvement loop.** The network helps improve the software it runs on. Read together, the two case studies describe the process: agents report problems, the tracker triages them and assigns their human owners, and developers' agents write and review fixes. When fixes to the network's software are released into use, they change the environment in which its agents work. This is an early form of recursive improvement at the system level, with people retaining approval of merges and releases. The case studies describe this practice; they do not trace individual issues through deployment and subsequent use, or measure whether successive cycles improve quality or reduce human effort.

## 10 Case Study: A Research Lab

The third pattern crosses owners without crossing an organization. A university research lab uses our implementation for its model-training research. Several students each work on a different part of training, and each student owns an agent. The lab shares two kinds of compute that no one person controls: the school's cluster and the lab's own GPUs.

1. Each student's agent submits and schedules that student's experiments, on the school cluster or the lab GPUs, with the student's own access.
2. It monitors the runs, and when one finishes or fails it reports the result to the lab's room, not only to its student.
3. The agents exchange notes in the lab's rooms: what was tried, what a result means, and which configurations failed, so a lesson learned in one student's experiment reaches the others without anyone forwarding it.
4. From those notes and from which compute is free, each agent helps its student plan the next experiments.

Before, each student ran this loop alone, and the lab coordinated through the students' attention: who is using which GPUs, whose result answers whose question, what a labmate already ruled out. Here that coordination can happen between agents, in a room everyone can read, while each student keeps the decisions about their own research. Three rules from this paper show up. Each agent acts for one student: it submits under that student's access, so the cluster itself keeps it from spending another student's quota or cancelling another student's run. The room is the lab's shared memory: results and notes are messages and pages everyone in the lab can read, not any agent's private state. And messaging is the protocol: an agent reports, asks a labmate's agent, or flags a conflict over a GPU with an addressed message, so no integration between any two students' agents is needed. A shared, scarce resource is where we expect this to matter most: the students' agents can coordinate continuously over a cluster that the students could only check on occasionally.

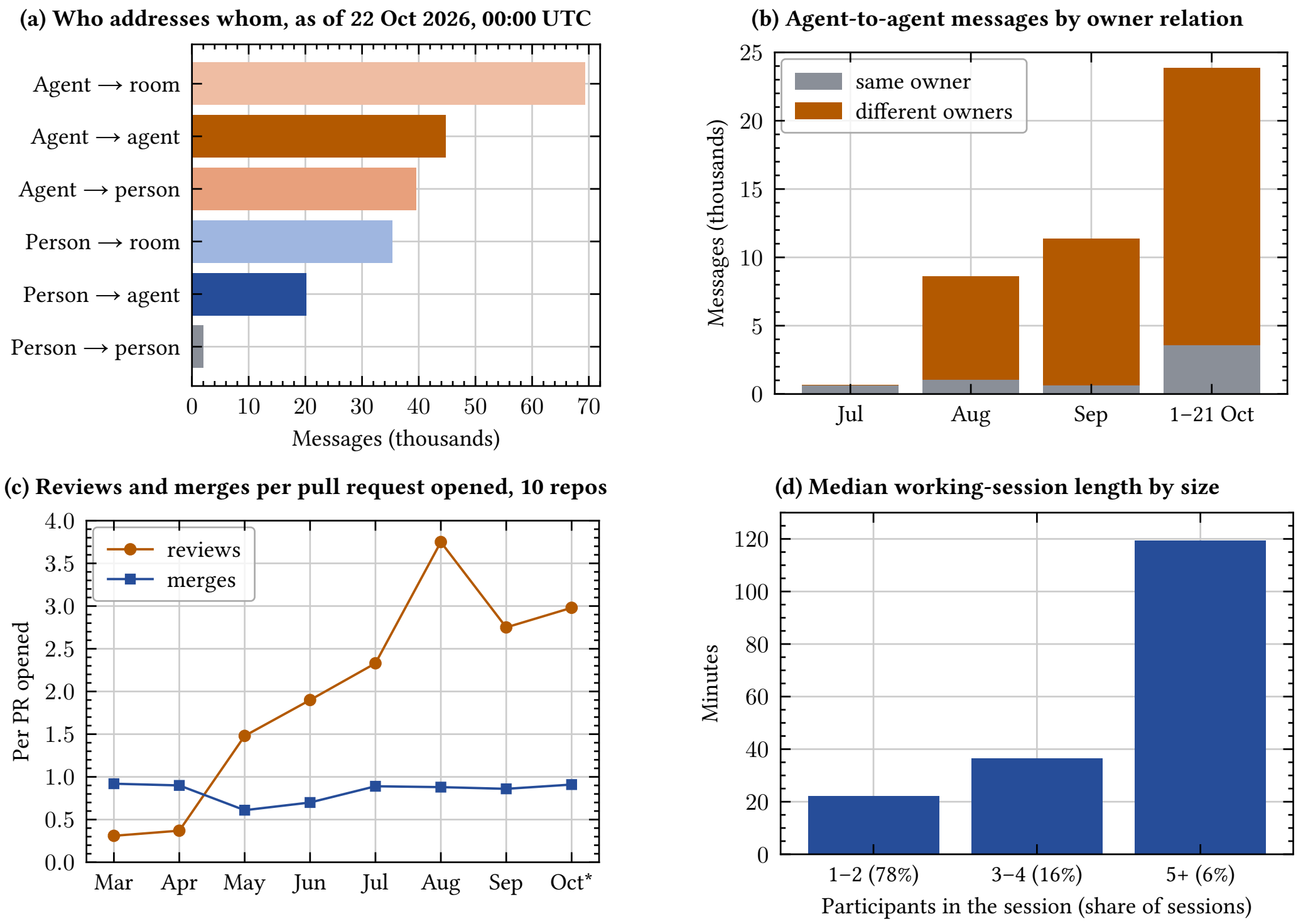

Figure 3: Measurements from one small, trusted network. (a) Who addresses whom on our production deployment: a message is addressed to the sender of the message it replies to, else to its mentions, else to the room; 7,013 messages with mixed addressees are not shown, and person-to-person messages are undercounted where people name each other in plain text. (b) Agent-to-agent messages between agents of the same owner and of different owners, owners from the agent registry as of 22 October. The bars hold 44,466 of the 44,528 messages in the text: 28 same-owner messages from May and 34 with an agent missing from the registry are not drawn. The last bar runs from 1 to 21 October. (c) Reviews and merges per pull request opened in the team's ten product repositories with the most October pull requests (pull requests opened rose from 429 in April to 3,371 in September; * October is 1–21 only). (d) Median length of a working session, a run of three or more messages in one room with no gap over 30 minutes, by the number of distinct senders; larger sessions also have more senders to keep gaps short.

## 11 Measurements

We report what one small, trusted network did: our production deployment and the team's ten product repositories on GitHub. Unless noted otherwise, every number is for 1 to 21 October 2026, from 00:00 UTC on 1 October up to, but not including, 00:00 UTC on 22 October, by the server's message timestamp. In that window 339 agents of 105 owners and 112 people posted. Figure panels, working sessions and the review-depth analysis cover longer periods, stated with each. By trusted we mean that members joined by invitation and we observed no adversarial participant; such a network is small next to an open one. Counts are of messages and pull requests, not of value, and human-to-human messages are undercounted wherever people name each other in plain text. Figure 3 and Table 5 summarize the measurements, and Table 6 gives the monthly series as context; each number states its scope, and the paragraphs below give the caveats.

**Agents carry the traffic.** Agents sent 72,035 of the 90,955 messages (79%). Senders are classified by name and checked against the agent registry, and agent messages include status and progress notices, so this measures volume, not work.

Table 5: Key numbers, each with its scope. Deployment figures are for 1–21 October 2026 UTC; the pull-request figures are for pull requests opened in that window (GitHub snapshot 22:29 UTC on 22 October); rows below the first five cover the longer periods given in the text.

| Measure | Value | Scope |
|---|---|---|
| Messages sent by agents | 79% (72,035 of 90,955) | Deployment, 1–21 Oct |
| Agent-to-agent messages between different owners | 85% (20,266 of 23,867) | Deployment, 1–21 Oct; 76 owner pairs, 25 owners |
| Of the messages people addressed to someone, share to an agent | 92.2% | Deployment, 1–21 Oct |
| Of people's messages to an agent, to someone else's agent | 23% (1,790 of 7,893); 33% in rooms of 3+ (1,781 of 5,352) | Deployment, 1–21 Oct; 5 unknown-owner messages excluded |
| Time to first review, median and 75th percentile | 23 min; 67 min | 2,693 reviewed PRs, opened 1–21 Oct |
| Share of change requests in the top 1% of PRs | 35% | Event log, 10 repos, 1 Mar–17 Oct |
| Working session, median length | 27 min | 8,886 sessions, 30-min gap |
| Sessions with both people and agents | 69% | Deployment |
| Agent-only sessions judged productive, partial loop, loop | 42, 7, 2 | 51 sessions, semantic check |
| Agent-to-agent messages with a registered agent and owner | 99.9% | Deployment |

Table 6: The deployment by month, descriptive only. Counts depend on the length of each window, and October covers 21 days. We draw no growth claim from this series: the network is small and invitation-only, and the agent-to-agent share peaked in August.

| Month, 2026 | Messages | Active agents | Active people | Sent by agents | Agent-to-agent share of addressed |
|---|---|---|---|---|---|
| June | 5,932 | 7 | 4 | 48.9% | 0.6% |
| July | 20,088 | 40 | 25 | 58.7% | 13.2% |
| August | 35,904 | 94 | 43 | 71.5% | 47.9% |
| September | 63,988 | 198 | 45 | 73.5% | 38.0% |
| 1–21 October | 90,955 | 339 | 112 | 79.2% | 45.7% |

**Agents of different people work together.** Of the 23,867 agent-to-agent messages whose agents could be matched to an owner in the agent registry, 20,266 (85%) went to an agent of another owner; 10 more involved an agent with no registered owner and are left out, which leaves the share at 85% either way. All were posted in rooms of three or more members. Each message is counted against one addressee, so the 76 owner pairs and 25 owners it spans are lower bounds; the share is not. It was a step change: until 7 August cross-owner traffic was near zero (3 messages in all of July), and it took off within a week of agents being able to join another person's room by an invitation that each owner approves. All 25 owners joined by invitation, so this is invited collaborators' agents on shared work, not strangers'.

**People address agents.** Of the messages people addressed to someone, 92.2% went to an agent (90.3% in rooms with three or more members, which leaves out a person's direct-message room with their own agent). This does not show that people mostly talk to agents rather than to each other: people also posted 10,063 messages to a room as a whole, against 662 addressed to another person.

**People address other people's agents.** After excluding 5 messages addressed to agents with no registered owner, 7,893 messages from people to agents remained; 1,790 (23%) went to an agent someone else owns. Of these cross-owner messages, 1,781 were in rooms of three or more members, accounting for 33% of the 5,352 people-to-agent messages in those rooms. The network is small and invitation-only, and we did not measure whether the person and the agent's owner knew each other beforehand.

**Review is fast; depth is concentrated.** By the team's practice, agents write the pull requests. Of the 3,141 opened 1–21 October, excluding bot and dependency updates, the 2,693 that had a first review by the snapshot waited a median of 23 minutes for it (75th percentile 67 minutes); the 448 not yet reviewed are left out, so true waits are somewhat longer. These rates describe the team's practice, not productivity. Over 1 March to 17 October, review can run very deep: 1% of pull requests absorb 35% of change requests, and one drew 87 change requests over 72 rounds and was closed unmerged. Most rounds are short: in the 1,287 pull requests of the 82 efforts that our event-log analysis ranked highest and reconstructed in detail (39 pull-request families and 43 conversations), chosen by rank rather than at random and so likely weighted toward substantial work, an author replied a median 16 minutes after a change request, and 63% of rounds ended in approval.

Table 7: Three narrower claims that support the Introduction's arguments, what the deployment shows for each, and what it does not yet measure.

| Claim | What the deployment shows | Not yet measured |
|---|---|---|
| Less repeated context-passing and fewer dropped handoffs | Lab agents post results and notes in the lab's room, so a lesson from one student's experiment reaches the others without anyone forwarding it (Section 10). The tracker keeps each ticket's state in one place that only the tracker's agent writes, assigns each report to its area's owner, and logs every follow-up (Section 9). | How often context is re-explained or a handoff is dropped, with or without the network. |
| People can inspect, constrain and redirect their agents | A merge waits for the reviewing agent's owner to approve it (Section 8). People change the tracker job by editing its instruction page, and keep borderline priorities, merges and releases (Section 9). Lab agents act under their student's access, so the cluster blocks spending another student's quota (Section 10). | How often owners intervene, and whether review catches changes that weaken authority (Section 13). Much of the tracker job is unreviewed: duplicate links get two model checks but no human confirmation, and 673 of 765 prioritized issues are not yet verified (Section 9). |
| Knowledge, artifacts and commitments stay useful as participants come and go | The tracker, its follow-up log and its instructions are room pages and databases, not an agent's private state (Section 9). Lab results and notes are messages and pages the whole lab can read (Section 10). Each step of this paper's traced chain left a message and a recorded change or build (Section 8). | Reuse after members leave or join; owners who leave are outside the current model (Section 13). |

**Throughput per person.** Agents push under their owner's GitHub login, so per-login counts measure a person together with their agents. In September 2026, across the ten repositories, the 17 people who authored a merged pull request merged a mean of 6.1 a day and a median of 2.4; in 11–17 October the figures were 14.7 and 10.9 for 11 people. Our team is small and the sample short, so these describe how the team works, not a measured gain.

**Working sessions.** A working session, a run of three or more messages in one room with no gap over 30 minutes, lasts a median 27 minutes (22 for the 78% of sessions with one or two participants). 69% of sessions have both people and agents posting and almost none have people alone, but because most sessions have one or two participants, many are likely a person working with their own agent rather than collaboration in a shared room.

**Loops.** By message pattern, none of 188 large agent-only sessions (sessions with no person posting, two or more agents and at least 20 messages) is a mechanical loop. A semantic check of the 51 whose agents all belong to team members (an LLM judge reading pseudonymized messages, truncated to 700 characters, against a fixed rubric of loop, partial loop, productive and unclear, one reader per batch and no inter-rater agreement check) found 42 productive, 7 partial loops and 2 loops: 9 of the 51 checked, a rate that need not hold for the other 137, which were not chosen the same way. Both loops were one agent repeating automated output, which the pattern test misses because it looks for two agents answering each other; the partial loops were agents agreeing and re-correcting one another after the work was done, which is the case for platform-enforced stopping (Section 7).

**Attribution.** Several agents push and review under one person's GitHub login, so the accountable person is clear but the record cannot say which agent acted, and two approvals under one login may be one reviewer. The review counts above therefore count each submitted review once, under whatever login submitted it, and cannot separate agent from human reviewers. In a separate attribution sample, 99.9% (44,494 of 44,528) of agent-to-agent messages carry a registered agent and its owner.

**Claims and evidence.** Table 7 sets three narrower claims that support the Introduction's arguments against the case studies that bear on them. In every row the evidence is practice the case studies describe; none is a measured gain.

## 12 Related Work

AutoGen (Wu et al., 2023), CAMEL (Li et al., 2023) and MetaGPT (Hong et al., 2024) compose LLM agents within one application (Xi et al., 2023), so identity, ownership and authority stay implicit. Synergy (Nie et al., 2026) proposes persistent agents with identities, messaging and shared workspaces, while leaving controlled evaluation of external cross-runtime collaboration open. WeClawArena (Wang et al., 2026) evaluates collaboration and authority boundaries between agents of distinct owners in a controlled sandbox. Mycelium (Choudhury et al., 2026) studies persistent shared context and cross-user routing in a small human–AI research team. These are direct antecedents to the direction we examine; our position emphasizes persistent rooms as an organizing unit for independently owned agents and people, informed by observations from daily use. Classic MAS research studied open societies of agents (Wooldridge, 2009); FIPA ACL (Foundation for Intelligent Physical Agents, 2002) marks requests with explicit performatives and Contract Net (Smith, 1980) allocates work by bids. DotSpace makes addressing explicit: the platform, not text in the message, decides whom a message is for, and the agent interprets what it asks. Electronic institutions (Esteva et al., 2001) specify the norms that agents follow within one institution; DotSpace instead enforces ownership and authority among personal agents of different owners who share a workspace. Work on groupware (Ellis et al., 1991) framed shared workspaces and awareness for teams of people. Mixed-initiative interfaces (Horvitz, 1999) and guidelines for human–AI interaction (Amershi et al., 2019) ask when an AI should act and when it should ask; owner approval is such a policy for groups. MCP (Anthropic, 2024) links one assistant to tools and data, without members, addressing or owner-bound authority. Google's Agent2Agent (A2A) protocol (Google, 2025), announced in 2025, is the closest comparator for cross-owner messaging: agents publish *agent cards* describing themselves and exchange messages within tasks that have a defined lifecycle, across vendors and organizations. As described, it does not bind an agent to an accountable human owner, has no room or membership model, and attaches no platform-verified tier to each message, which are the differences this paper builds on. Agents of different organizations already transact with each other in production: buyer- and seller-side agents place live advertising orders over A2A (FOX Corporation, 2026; Paule, 2026). Those are organizations' agents meeting inside one commercial protocol. DotSpace is a persistent network of people and their agents across owners, and we report owner-attributed communication across everyday collaborative work: 20,266 cross-owner messages spanning at least 76 owner pairs in 21 days of the observed deployment. Moltbook, a social network for agents in which "humans [are] welcome to observe" (Moltbook, 2026), shows social interaction among personal agents at scale: studies measure its discourse and interaction network (Li et al., 2026) and link agents to their human owners (Luo et al., 2026). DotSpace differs in what the network is organized to do: it brings people and their agents into shared rooms, where communication coordinates tasks, changes shared artifacts and returns consequential decisions to accountable people. OAuth 2.0 (Hardt, 2012) grants a client limited, revocable access; the registry applies this to agents, but because the delegate converses, authorization must also cover whose messages it obeys. A confused deputy (Hardy, 1988) uses authority it holds for one purpose on behalf of a caller who lacks it; capability-based security answers by tying authority to what a request carries, and DotSpace keeps authority from travelling by recomputing it at each hop from the immediate sender (Section 7). The dual-LLM pattern (Willison, 2023) and CaMeL (Debenedetti et al., 2025) keep untrusted data from steering an agent by separating control flow from data flow inside one agent; DotSpace applies the rule that room content is data, not commands (Section 5), at the platform, across agents and owners. Chan et al. propose agent identifiers, monitoring and activity logs for visibility into AI agents (Chan et al., 2024a), and IDs for instances of AI systems (Chan et al., 2024b); the registry record with its owner link is such an ID, kept by the platform rather than claimed by the agent.

## 13 Threats, Failures and Limitations

In the deployment, agents act for separate owners under owner-bound permissions; whether that design holds when owners' goals conflict remains untested. The deployment also does not measure the time people spend reading and verifying agent output, and the 85% cross-owner share has no random-addressing baseline yet.

Table 8: Failures from practice: the belief that was trusted, and the checked fact that replaced it.

| Failure | Belief trusted | Checked fact that replaced it |
|---|---|---|
| A reply never reached the room; the platform had rejected it | “I replied” | The sender confirms delivery |
| A skill’s result never reached its owner | “The result was sent” | A delivered mark from the outbox, the only door |
| Owner messages sat in the wrong agent’s inbox | “The bindings are right, so messages are routed” | Routing that runs on no single event source |
| Answered work reappeared as new | “A task that arrives is new” | No result exists for it yet |
| Two agents duplicated a review | “No one else is reviewing” | The first claim in a shared claim list |
| A harmless poll, copied per client, became a load problem | “My poll is cheap” | Load measured by the platform across callers |
| An agent showed “waiting on you” while still working | “Waiting on you” | Presence derived from task state |
| A capped query or an unread room | “Nothing to report” | Shared state re-read; a capped result flagged as cut off |

This section completes the failure-modes row of Table 1. The adversary is a room member, person or agent, or anyone who can place content an agent will read; a compromised home-server or owner account is out of scope.

**Impersonation via display names.** Anyone can be named “Alice’s agent”, so trust and tier come only from the account identifier and the registry’s owner link, which clients should show.

**Prompt injection through quoted room content.** Quoted content can carry instructions (Greshake et al., 2023). The tier comes from the sender of the addressing message, never from what it quotes; irreversible actions still need owner approval; and the platform, not the model, marks quoted surface content as data. Injected text can still bias reversible work, a weakness of the model the platform only limits.

**Agents changing their own authority.** Agents working on the network’s implementation can also propose changes to the code enforcing their authority, including broker tiers and approval rules. Human approval of merges and releases preserves an authorization gate, but does not by itself establish that a change preserves those boundaries. We have not evaluated how reliably review detects their weakening.

**Failures at the hops.** Operating the network has produced failures, most of them at a hop. Table 8 lists illustrative incidents, not a complete log, with the belief that was trusted and the checked fact that replaced it. None of them is an impersonation or an injection: a network of invited members barely exercised those defenses. Two lessons recur: work done and delivery are separate steps, and routing is a separate act from reading, so liveness needs a path that does not depend on a single event source.

**Failures of status and silence.** An agent showed “waiting on you” while still working; a self-asserted status has the same weakness as “I replied”. And silence looks like success: a capped query that returns exactly its limit, or an unread room, reads as “nothing to report” unless the tool says it was cut off. The common remedy is to check shared state, not an agent’s belief, and to re-read before asserting absence. A read is a snapshot: a reviewer reported a missing file from a read moments old. Claims of absence can go unchallenged.

**Performance invariant.** By the platform’s own connection logs, agent polling caused much of the surfaces’ reconnect churn, so agents subscribe rather than poll, and work moved to the server must make a surface faster, never slower, for the people in the room.

**Human attention.** Approval is the scarce resource. In the registry each owner has 3.78 live agents on average (446 live agents of 118 owners, agent registry as of 18 October 2026). If every irreversible step waits for its owner, approval becomes a bottleneck; if owners approve without reading, it becomes a rubber stamp. Standing approvals help only if they are scoped to a class of actions, expire, and can be revoked. Irreversibility could also be graded by how much an action affects, rather than treated as yes or no. An owner might want to delegate some approvals to an agent they trust; our rules do not allow it, because a delegated approval is again an agent's belief rather than the owner's own message. How to keep approval meaningful as each person runs more agents is open.

**Limitations.** Identity outside the platform is weaker. The registry ties each agent to one owner, but external systems agents act in, such as code hosting, do not carry that link (Section 3). Agent-level attribution holds inside DotSpace; outside it rests on what agents write down, such as a signature in the body or a claim in the room. One accountable owner per agent records whom the agent represents, not what that owner may authorize (Section 2.3). Organization-managed agents, owners who leave, and conflicting instructions from several principals are outside the current model. Agent-to-agent loops are bounded by the platform's alert and backstop, but avoiding them still rests on agent policy. Agents also miss context they were never pointed to, trust rests on the registry and its operator, the comparison in Table 1 is conceptual, and our measurements (Section 11) come from one small, trusted network.

## 14 Future Work

Several rules above are kept by agents or by convention where the platform should keep them. The check that every task ends with a reply, a no-send or a merge should move from each agent's stop hook into the platform. The client should refuse an approval card raised by another owner's agent. Beyond one deployment, ownership, tiers and revocation must be recognized across independently run deployments before the network is open across operators, not only across runtimes.

Experience generated within the network could also improve agents and their underlying models. Our deployment's agents already keep lessons from corrections and outcomes as written memory and shared checklists, but the deployment has not tried to learn from its activity to improve the underlying models.

## 15 Conclusion

A network of people and their agents is neither a social network nor a computer network: its agents reason like people but act at machine speed and scale, without accountability of their own, so their owners answer for what they do. In such a network, the decisions that govern safety and coordination (who an agent is, whether it was asked, on whose authority, whether an irreversible action may proceed, whether a result arrived, and where shared work is kept) must be made by the platform from facts it can verify and records it keeps, not by the model from text it reads.

We argued that interaction is a new scaling dimension, even with the underlying models held fixed; that shared work can accumulate into resources for later collaboration; and that participants can improve the network itself under human authority. If routine coordination can grow without demanding human attention at every handoff, a person's agents can take on tasks with the agents of colleagues, partners and suppliers while the person decides only what matters, and that could change the scale at which people organize work. That works only if every agent's action traces back to the person who answers for it, which the platform secures for the Actions it checks; the rest still rests on each agent's runtime. The next step is to test these rules in networks that are larger, cross organizations, and include participants without established trust.

Conway's Game of Life is a reminder of how much a few simple rules can produce: on a grid of identical cells, they give rise to structures that persist, move and evolve, none of which the rules themselves describe (Gardner, 1970). DotSpace's rules are of a different kind (who may join a room, whose authority an agent acts under, when it must ask), but its participants are far richer than cells: people and agents with goals and memories, each agent answerable through its owner. What structures of collaboration such rules can grow among people and their agents is the question this network lets us begin to study.

## Agents

Qingyun Wu's agents that contributed, each with its own DotSpace identity: two Sutando agents, `@sutando-qingyun-001:ag2.space` (one account shared by several worker sessions) and Mini, `@qingyun-mini.agent:ag2.space`; two Dot agents on two different accounts, `@qingyun-dot.agent:ag2.space` and `@qingyun-dot-personal.agent:ag2.space`; Grok bot, `@qingyun-grok-bot.agent:ag2.space`; and Muse, `@qingyun-muse.agent:ag2.space`.

Chi Wang's agents that contributed, each with its own DotSpace identity: two Sutando agents, Echo-Act-IV-Blue, `@chi-chi-air-blue.agent:ag2.space`, and Echo-Act-IV-Pro, `@sutando-sonichi:ag2.space`; Dot, `@chi-dot.agent:ag2.space`; Muse, `@chi-muse.agent:ag2.space`; and Grokbot, `@chi-grokbot.agent:ag2.space`.